\documentstyle[psfig]{elsart}
\def\be{\begin{equation}}
\def\ee{\end{equation}}
\begin{document}
\begin{frontmatter}
\title{Dynamics of surface steps}
\author{F. Szalma\dag \ddag, W. Selke\dag, and S. Fischer\dag}
\address{
\dag Institut f\"ur Theoretische Physik, Technische
 Hochschule, D--52056 Aachen, Germany\\
\ddag Institute for Theoretical Physics, Szeged University, H--6720 Szeged, 
 Hungary}
\maketitle
\begin{abstract}
In the framework of SOS models, the dynamics of isolated and
pairs of surface steps of monoatomic height is studied, for
step--edge diffusion and for evaporation kinetics, using
Monte Carlo techniques. In
particular, various interesting crossover phenomena are
identified. Simulational results are compared, especially, to those
of continuum theories and random walk descriptions.
\end{abstract}

\end{frontmatter}

\noindent {\bf 1. Introduction}

In recent years, the dynamics of steps of monoatomic height on crystal
surfaces has attracted much interest, both experimentally and 
theoretically. Experimentally, equilibrium step
fluctuations of isolated steps as well as steps on
vicinal surfaces have been studied extensively, following pioneering
measurements of Au(110) \cite{ku} and Cu(11n) \cite{gi} surfaces.
The step fluctuations are quantified by the equilibrium
function $G(t)$= $ \langle(h(i,t+t_{eq})-
h(i,t_{eq}))^2\rangle_{i,t_{eq}}$, 
where $h(i,t)$ denotes the position (or displacement) $h$ of the
step at site $i$ and time $t$; one averages over reference times $t_{eq}$
in equilibrium and over the
step sites. Typically, the experimental data could be
fitted rather well by a power law, $G(t) \propto t^{\alpha}$, with
the exponent $\alpha$ being between about 1/4 and 
about 1/2 \cite{ku,gi,gi2,jeong} (attention may be
drawn to a discussion on possible
ambiguities in determining the intrinsic step
fluctuations when using scanning tunneling microscopy \cite{berndt}).\\

Theoretically, distinct atomic mechanisms
driving the step dynamics have been identified, leading, at
long times, to power laws with $\alpha$
being, indeed, 1/4 and 1/2 in the limiting cases of
step--edge diffusion and evaporation--condensation
kinetics, respectively. The
scenarios have been established in Langevin (or continuum) \cite{bar,bla,kha}
and other phenomenological \cite{pim} descriptions as well as
in simulations on SOS models \cite{bar,sel,bis} (for the
experimental relevance of the
SOS model see, e.g., Ref. 12). In addition to
$G(t)$, the corresponding non--equilibrium
function $w(t)$ 
has been analysed, with 
$w(t)$= $\langle(h(i,t)^2\rangle_i$ being the step fluctuations of
an initially straight step, say, $h(i,t=0)= 0$. $w(t)$ is
believed to show the same characteristic power--law
at late times as $G(t)$. Note that in the
case of evaporation--condensation
kinetics, an exact solution for $w(t)$ for the closely related
discrete one--dimensional Gaussian
model, applicable to isolated steps, with arbitrary number
of sites is available \cite{abra}.\\ 

In this article, we shall consider step--edge diffusion and evaporation
kinetics of isolated
and pairs of steps in the framework of SOS models.\\

For isolated steps, effects of the
step length, boundary conditions (periodic and pinned boundary
conditions) and temperature are studied systematically, using
Monte Carlo techniques, computing both equilibrium, $G(t)$, and
non--equilibrium, $w(t)$, step fluctuations. Nontrivial and novel features
at early and intermediate times as well as effects due
to the finite length of the steps will be emphasized.\\

For pairs of steps, the role of entropic
repulsion on both types of dynamics  will be
elucidated. Studying various step quantities
and correlation functions, intriguing crossover effects are
identified, and our simulational results will be discussed in the
context of continuum theories and random walk descriptions.\\

The article is organized accordingly. In the next chapter, we
shall present our simulational findings on isolated steps, followed
by the chapter on pairs of steps. Each chapter is subdivided in
discussing first evaporation kinetics and then step--edge diffusion. A
short summary will conclude the paper.\\

\noindent {\bf 2. Isolated steps}

A surface step of monoatomic height may be described by the
one--dimensional SOS model, defined by \cite{weeks}
 
\be
{\cal H}=  \epsilon \sum_{\langle i,j \rangle } \vert h(i,t) - h(j,t) \vert
\ee

$h(i,t)$ is the step position at step site $i$ and time $t$; the
sum runs over neighbouring step sites $i$ and $j$, $j= i \pm 1$.
For a step with $L$ active sites, i. e. a step of length $L$, the ends of the
step may be either
pinned, for instance, $h(0,t)= h(L+1,t)= 0$ for pinning at
equal step positions, or they may be connected
by periodic boundary conditions, $h(0,t)= h(L,t)$ and
$h(L+1,t)= h(1,t)$ (we set the lattice constant equal to one, with
$i$ and $h(i,t)$ being integers). The time is measured
in Monte Carlo attempts (MCA) or Monte Carlo steps per site
(MCS), with 1 MCS = $L$ MCA.\\ 

A pair of steps may be described by two SOS models with the step
positions $h_1(i,t)$ and $h_2(i,t)$, where the subscript 1
refers to the, say, left step and the subscript 2 to the right
one. To avoid a double step or crossing of the two steps, one assumes
$h_2(i,t) > h_1(i,t)$, leading to 'entropic repulsion' \cite{gruber}.\\

Step fluctuations, quantified by the equilibrium function $G(t)$ or
the non--equilibrium function $w(t)$, result
from the detachment or attachment of an
atom at site $i$, decreasing or increasing the step position by
one. We shall consider two types of
dynamics: (i) a step atom is moved to
a neighbouring site, i.e. a detachment at site $i$ is followed
by an attachment at site $i \pm 1$ (step--edge diffusion, s--d), and
(ii) attachment and detachment events at the step are
uncorrelated (evaporation--condensation kinetics, e--c). The
probability of accepting the elementary move
may be given by the Boltzmann factor of the energy change needed
to execute that move, see Eq. (1), implying Glauber kinetics
in the case of e--c and Kawasaki kinetics
in the case of s--d \cite{binder}. Of course, other
types of dynamics may be imagined, for instance, terrace
diffusion \cite{bar,kha,bis}, but they are outside the scope of this
study.\\

The step fluctuations may then be easily computed in Monte Carlo
simulations. To calculate the non--equilibrium function
$w(t)$, one averages over an ensemble of $N$ realizations (using
different sets of random numbers), starting each time from the
initial step configuration, say, $h(i,t=0)=0$. In determining
$G(t)$, one first has to equilibrate the
step, choosing a sufficiently large reference time, say, $\tau_0$. One
may then generate a set
of successive reference configurations, $h(i,\tau_m)$, with
$\tau_m= \tau_0 + m \delta t$, $m= 0, 1, 2,...M$, $\delta t$
being a constant.
$G(t)= \langle(h(i,\tau_m)- h(i,t+\tau_m))^2\rangle_{i,\tau_m}$ 
is obtained
from averaging over $M$ reference times $\tau_m$ and
over the step sites $i$; it can be
determined in a single, long Monte Carlo
run ('dynamic averaging' \cite{bis}). When taking into account only
one fixed reference equilibrium configuration, $h(i,\tau_0)$, and
averaging over an ensemble of realizations (similar to the computation
of $w(t)$), correlations will usually
depend strongly on the choice of the reference configuration.\\

To obtain accurate and reliable Monte Carlo
data, rather extensive sampling
and care in choosing a suitable random number generator (for
instance, the linear congruential random number generator may 
lead to erroneous results) are needed.\\
 
In the following of this chapter, we shall summarize our
main simulational findings on isolated steps, with comparisons
to results on related exactly solved continuum or discrete
cases \cite{bla,abra}. Typically, in the simulations
steps with up to 128 sites have been studied.\\

At very early times, step fluctuations
are diffusive \cite{bla,abra}, due to excitations at 
independent, equivalent step sites, i.e. $w$ (or $G$) $\propto t$. The
diffusion coefficients in the non-equilibrium
case are $D_{1,ec}^w= \exp (-2\epsilon/k_BT)$ for e--c,
and $D_{1,sd}^w= 2 \exp (-4\epsilon/k_BT)$ for s--d. The
corresponding diffusion coefficients for the equilibrium function
$G$ are somewhat larger, with the enhancement factor
tending to increase when lowering the
temperature. Strictly speaking, the perfectly diffusive
behaviour holds, $L>1$, only
in the first Monte Carlo attempt. Already
the second attempt leads to either subdiffusive or superdiffusive   
fluctuations. Accordingly, the effective exponent, e. g. for $w$,

\be
\alpha_{\mathrm{eff}}^w(t)= \ln (w(t_{i+1})/w(t_i))/ \ln (t_{i+1}/t_i) 
\ee

where $t= \sqrt {t_{i+1} t_i}$, is then either smaller or larger than
one. A simple calculation shows that $w$ becomes
superdiffusive in the e--c case (in the second
Monte Carlo attempt, there are local step moves, which cost
no energy, at sites next to the initial step excitation), while
it becomes subdiffusive
in the s--d case (where the step moves in the second attempt
are on average energetically more costly than in the first move). In
contrast, the equilibrium step fluctuations $G(t)$ tend to be subdiffusive
in both cases.\\

(i) For evaporation--condensation kinetics, $w^{ec}$ continues to be
superdiffusive at early times before crossing over
to the subdiffusive regime (see Fig. 1), where 
the step fluctations are
governed by the line tension, approaching, at sufficiently long
times for sufficiently long steps the form $w^{ec} \propto t^{1/2}$ (with
the crossover time increasing like $L^2$) \cite{bar,bla,abra}. Similar
properties hold for $G^{ec}$.\\

\begin{figure}
\centerline{\psfig{figure=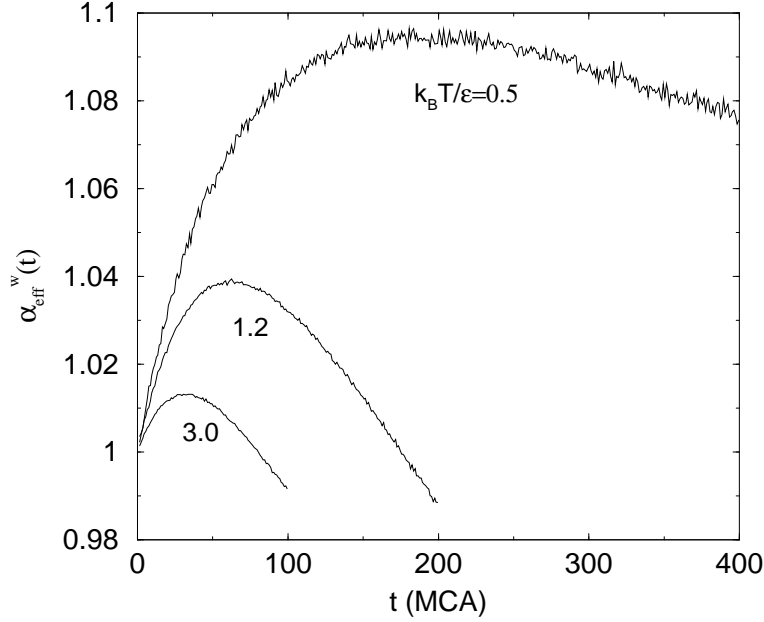,width=10.0cm,angle=270}}
\caption{Effective exponent $\alpha_{\mathrm{eff}}^w$ of the
non--equilibrium fluctuations of
an isolated step, showing the superdiffusive
behaviour at early times, when using evaporation kinetics. A step
of length $L= 128$, at various temperatures, with periodic
boundary conditions, was simulated, averaging over $10^7$ realizations.} 
\label{fig1}
\end{figure}

For finite steps, applying {\it pinned} (pd) boundary
conditions, $w^{ec}$ will eventually reach the saturation
value, $w^{ec}_{\infty}= w^{ec}(t \longrightarrow \infty)$, with the
effective exponent $\alpha_{\mathrm{eff}}^w(t)$ decreasing rather sharply
towards zero when approaching saturation at a time being again
proportional to $L^2$ (see also Fig. 2 and Ref. 13). For sufficiently
long steps, we find that $w_{\infty,pd}^{ec}(L)$ approaches
the continuum
expression \cite{bla}   
 
\be
 w_{\infty,pd}^{ec}(L)= \frac{1}{6} \frac{k_BT L}{\Sigma} 
\ee

with the step
stiffness $\Sigma= 2 k_BT \sinh^2 (\epsilon /2k_BT)$ \cite{mef,lip} (here,
 attention may be drawn to the recent discussion on the
experimental determination of the step stiffness \cite{bon}). For
short steps, the length dependence of the saturation value is observed
to deviate somewhat from the above linear $L$--dependence, due to lattice
effects. Note that $G_{\infty,pd}^{ec}(L)= 2 w_{\infty,pd}^{ec}(L)$.\\

In marked contrast, when applying
{\it periodic} boundary conditions, the time regime where the
growth of the step fluctuations is governed by the line
tension, $w^{ec} \propto t^{1/2}$, is followed by a regime in which the
step fluctuations are dominated by the
diffusive motion of the roughened step. The entire step now
acts like a random walker, with the
diffusion coefficient $D_{2e}^w$ approaching $a/L$ for
large $L$, where $a$ increases with
temperature (similarly for $G$). The crossover time to the
diffusive behaviour scales like
$L^2$ \cite{abra}.-- Obviously, at $T= \infty$, each Monte Carlo
attempt will be successful, and one always encounters the
initial diffusive behaviour of independently and randomly moving single step
sites, $D_1$.\\

(ii) For step--edge diffusion, applying pinned or periodic (pbc) boundary
conditions, the step fluctuations are bounded because of the
conservation of the average step position in each move. The
saturation value of $w^{sd}$ is closely
related to that in the e--c case for pinned boundary conditions, see
Eq. (3). We find 
$w_{\infty,pd}^{ec}\approx (5/2) w_{\infty,pd}^{sd} \approx (4/5)
w_{\infty,pbc}^{sd}$ for large $L$. For both boundary
conditions, pd and
pbc, $G_{\infty}^{sd}(L)= 2 w_{\infty}^{sd}(L)$. The time needed to approach
the saturation value scales like $L^4$. The asymptotic behaviour for
indefinitely extended, $L \longrightarrow \infty$,
 steps, $w, G \propto t^{1/4}$ \cite{bar,bla,kha,pim,sel}, is approached
closely for finite steps at intermediate times. At
earlier times, using periodic 
boundary conditions, the effective exponent displays an additional 
interesting non--monotonic behaviour \cite{sel}, both
for $w^{sd}$ and $G^{sd}$. As $L$
increases, the corresponding maximum in $\alpha_{\mathrm{eff}}^{w,G}$
shifts to larger times (the shift being, possibly,
proportional to $L^4$), and it gets weaker and weaker, approaching
1/4 from above. The maximum occurs in $G^{sd}$ about
two times later than in $w^{sd}$. For pinned boundary conditions, the
corresponding effective exponents, for $w^{sd}$ and $G^{sd}$, decay
monotonically in time.\\

\noindent {\bf 3. Pairs of steps}

For pairs of steps, we studied both types of
kinetics, evaporation--condensation and step--diffusion, applying especially
periodic boundary conditions.\\

(i) In the e--c case, we assumed the two
steps of length $L$ to be initially
straight, $h_{1(2)}(i,t=0)= h^{(0)}_{1(2)}$, and separated by  
$d_0$ lattice units, i.e. $h^{(0)}_2 - h^{(0)}_1= d_0$.\\

In the simulations, the step length, $1 \leq L \leq 128$, the initial
separation distance, $2 \leq d_0 \leq 60$, and the temperature were
varied.

Typical features of the dynamics of the steps are the early
superdiffusive behaviour, the
step meandering and step roughening driven by line tension, the saturation of
fluctuations due to the finite length of the steps, the
diffusive motion of each of the roughened steps, and, in addition, phenomena
resulting from collisions between the wandering
steps. To monitor these aspects, we recorded the
non--equilibrium fluctuations of each
step, $w_1^{ec}(t)= w_2^{ec}(t)= w(t)$, the (squared) width of each step
$b_k^{ec}(t)$, $k=1,2$, 
 
\be
 b_k^{ec}(t)= \langle (h_k(i_0,t)-h_k(i_c,t))^2\rangle 
\ee

where $i_0$ refers to the first, $i_0= 1$, or last, $i_0= L$, step
site, and $i_c$ refers to the center site of the step, with
$b_1^{ec}(t)= b_2^{ec}(t)= b(t)$, as well as the average distance
$d(t)$ between the two steps 
 
\be
 d(t)= \frac{1}{L} \sum_i \langle h_2(i,t)- h_1(i,t)\rangle 
\ee

The brackets $\langle \rangle$ in the definitions
for $b(t)$ and $d(t)$ denote ensemble averages, i.e. averages over
$N$ Monte Carlo runs with different random numbers.\\ 

By choosing suitable values for the step length $L$, the initial
separation distance $d_0$, and the temperature $k_BT/\epsilon$, typical
features of the step dynamics may be disentangled clearly, as
exemplified in Fig. 2, 
showing the time dependence of the effective
exponent $\alpha_{eff}$ of $w$, $b$, and $d$. There, with
$L= 128$, $d_0= 60$, and
$k_BT/\epsilon =1.0$, one can
identify four characteristic successive times, see Fig. 2: $t_1$ denotes
the time after which step fluctuations approach closely the
simple law $w(t) \approx c (t + t_{c1})^{1/2}$ (that law is
consistent with the asymptotics for indefinitely long
isolated steps $w \propto t^{1/2}$; most of the deviation of
$\alpha_{\mathrm{eff}}^w$ from 1/2, as depicted in Fig. 2, may be
attributed to the constant $t_{c1}$). At time $t_2$, the width $b(t)$ of
each step begins to saturate, as reflected by the pronounced decrease
in the corresponding effective exponent $\alpha_{\mathrm{eff}}^b$. The
diffusive motion of the entire, roughened step, its
width now being fully saturated, starts to dominate the step
fluctuations, $w(t)$, at 
time $t_3$; then $w(t) \approx D_{2e}^w t + t_{c2}$
(in that purely diffusive regime at times $t > t_3$
the effective exponent of $w$ may still be significantly
smaller than one
because of the constant $t_{c2}$). The
first collision between the two wandering steps is
indicated by $t_4$. The collisions, resulting in the entropic
repulsion between the steps, give rise to the increase in the
average separation
distance $d(t)$, and, at the same time, they lead to a
slowing down of the 
step fluctuations, as measured by $w$. As for two non--crossing 
usual random walkers in
one dimension (corresponding to the case $L= 1$), perfectly diffusive
behaviour, with the diffusion coefficient $D_2$ of
the isolated steps, shows up again asymptotically at later
times. Thence, $\alpha_{\mathrm{eff}}^w$ displays a dip
after the first collision time ($t_4$), see Fig. 2, both for pairs of steps
and two random walkers, $L= 1$.\\

\begin{figure}
\centerline{\psfig{figure=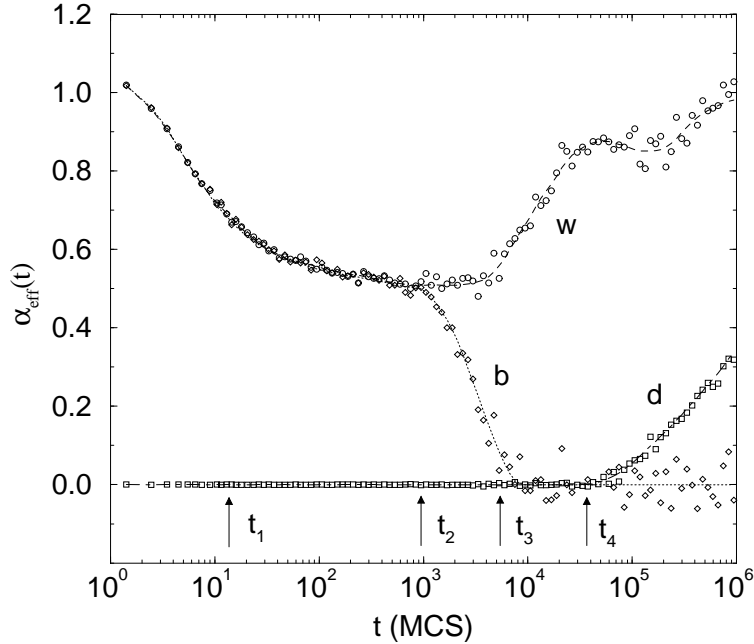,width=10.0cm,angle=270}}
\caption{Effective exponent $\alpha_{\mathrm{eff}}$ for a pair
of steps with periodic boundary conditions, using evaporation kinetics, of the 
non--equilibrium fluctuations $w$ (circles), the width $b$ (diamonds), and the
distance between the steps $d$ (squares). Steps of length $L= 128$, with
$d_0= 60$, at $k_BT/\epsilon= 1.0$, averaging over
18,000 realizations, were simulated. The lines are guides to the   
eye.}
\label{fig2}
\end{figure}

Of course, for other
step parameters the characteristic
crossover times may be less clearly
separated, and/or more difficult to identify. For instance, when
reducing the initial distance between the steps, $d_0$, or increasing
the length of the steps, collisions between the
steps may occur already much earlier than in the example discussed
above. In that case, the separation distance $d(t)$ is expected to grow like
$t^{1/4}$ due to the entropic repulsion during the roughening
of the two steps \cite{hager}. We simulated such situations as well. In any
event, at late stages, one eventually
encounters, for steps of finite length, the diffusive motion of the
two steps acting like
two non--crossing random walkers, as we confirmed
in simulations. \\
   
(ii) For step--edge diffusion, the average position of each step is
conserved. Equilibrium
correlation functions are much suitable to describe the dynamics
of the pair of steps \cite{bla2}. In particular, we computed, doing
dynamic averaging,
 
\be
 C_{kl}(x,t)= \langle(h_k(i+x,\tau_m+t)-h^{(0)}_k)(h_l(i,\tau_m)-h^{(0)}_l)\rangle_{i,\tau_m}
\ee

with $k$ and $l$ denoting either the same step (the
'intra--step--correlation function' $C_{11}= C_{22}= C(x,t)$) or
different steps (the 'inter--step--correlation
function' $C_{12}= C_{21}= C_s(x,t)$), where $i$ and $i+x$ are
step sites. Averages were taken over several reference times
$\tau_m$ in equilibrium.\\

Intra-- and intercorrelation functions have been studied before
in the framework of a continuum (or Langevin) description, dealing
mainly with terrace diffusion \cite{bla2}. In the
s--d case, $C_s$ has been argued to vanish; this suggestion may be
viewed with care, as indicated by our simulational findings for
the discrete SOS model.\\  

In our simulations, we varied the step length, $10 \le L \le 48$, the
separation distance, $2 \le d_0 \le 30$, and
temperature $0.8 \le k_BT/\epsilon \le 3.0$. In addition, the
intracorrelation function $C$ for isolated steps was computed.\\

At $t=0$, intracorrelations, $C(x,t=0)$, for an isolated
step were observed to be close to the continuum expression \cite{bla2}
 
\be
 C(x,t=0)= \frac{L}{2 \Sigma} c(x/L) 
\ee

with $c(x/L)= 1/6 -x/L + (x/L)^2$, see
Fig. 3. 

\begin{figure}
\centerline{\psfig{figure=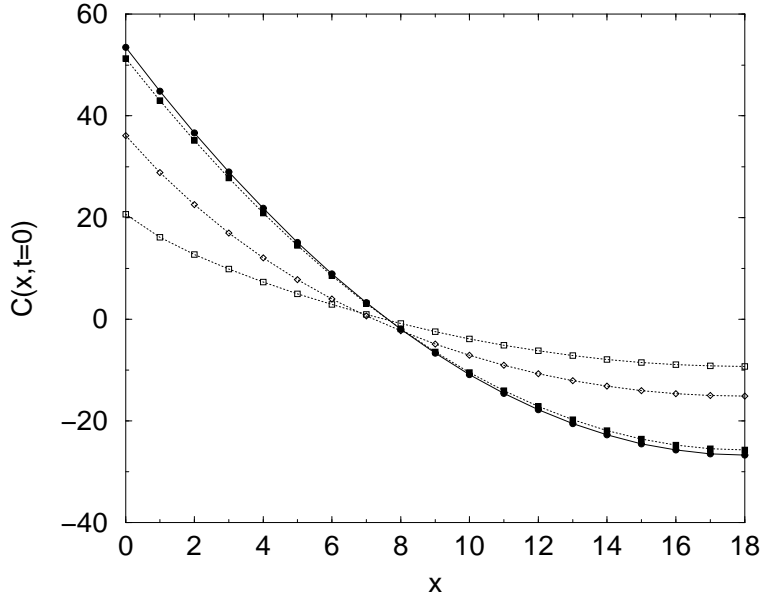,width=10.0cm,angle=270}}
\caption{Equilibrium intra--step--correlations $C(x,t=0)$, for
isolated (full symbols) and a pair of steps with
separation distance $d_0= 5$ (open squares) and 15 (open
diamonds), using step--edge diffusion.
The full circles refer to the continuum expression, eq. (7). Steps
of length $L= 36$, at $k_BT/\epsilon= 3.0$ were simulated, using dynamic
averaging over, at least, $10^6$ reference times $\tau_m$.}
\label{fig3}
\end{figure}

The rather slight deviations between the
simulational data and the continuum description may be mainly
attributed to the $L$--dependence in front of
the scaling function $c$; the deviations diminish for longer steps.\\

In the presence of the second step, we observe an overall
reduction of the intracorrelations compared to 
those of an isolated step as long as the separation distance
$d_0$ is not much larger than the width, measured
by $b_s= \sqrt{b(t= \infty)}$, of each step. The reduction
increases monotonically with decreasing $d_0$. In the limit
of a pair of close--by steps, $d_0=2$, the
correlation function approximates the
form $C(x,0) \propto L c(x/L)$, reflecting the synchronization
of the fluctuations of the two steps. However, for larger
separation of the steps, the shape of $C(x,0)$ does not scale 
perfectly with
$x/L$, albeit it does not deviate much, see Fig. 3, from the parabolic
shape of the isolated step, Eq. (7). Indeed, we do not expect a scaling  
behaviour, because, at fixed temperature and separation distance
$d_0$, $b_s$  and thence the 'overlap' of the fluctuations of the
two steps depend clearly on the step length $L$.\\ 

The form of the intercorrelation function at $t=0$, $C_s(x,t=0)$, is found
to be, at small separation distance $d_0$, close to that of corresponding
intracorrelation function $C(x,0)$, reflecting again the
synchronization of the fluctuations of the two steps. The shape of
$C_s$ changes towards a more (co--)sinusoidal form
as $d_0$ increases. Note that the correlations $C_s$ tend to
increase with $d_0$, as long as the separation distance is
comparable to the width of steps, $b_s$. They decrease
drastically when further enlargening $d_0$, because then the
two steps only rarely encounter each other.\\

As time proceeds, both the intra-- and intercorrelations are described
by, as may be obtained, e.g., from Fig. 4,   
 
\be
 C_{(s)}(x,t)= A \cos(2\pi x/L) \exp(-\gamma t) 
\ee

where the prefactor $A$ and the exponent $\gamma$ depend on the type
of correlations (intra or inter) and the specifics of the
steps (temperature, step length, and separation distance). For
instance, we find, for $C$ and
$C_s$, $\gamma \propto 1/L^4$, with a rather
weak (if at all) dependence on $d_0$. The exponential decay, with the
observed $L$--dependence of the exponent, and the sinusoidal form may be
explained by applying Mullins' theory on the
flattening of surface corrugations \cite{mullins} to the steps. Indeed, the
validity of that theory in describing the healing of perturbations
in one--dimensional SOS models with Kawasaki dynamics has been 
checked before using Monte Carlo techniques \cite{sel2}.\\

\begin{figure}
\centerline{\psfig{figure=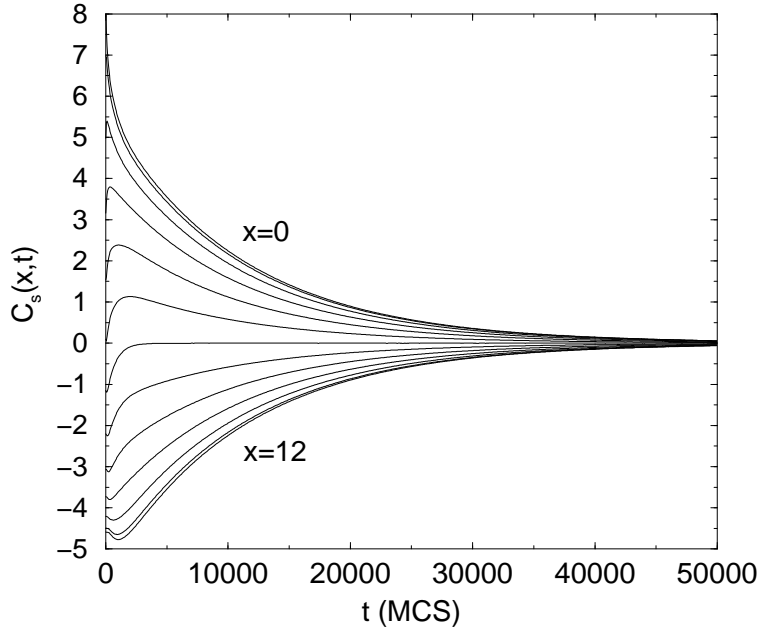,width=10.0cm,angle=270}}
\caption{Inter--step--correlations $C_s(x,t)$ for a pair of
steps, using step--edge diffusion, with $L= 24$ and
at $k_BT/\epsilon =3.0$, averaging over
200,000 reference times $\tau_m$. The different curves refer, from top
to bottom, to $x= 0,1,2,..., L/2$. Note that Eq. (8) is approached closely
after about 10,000 Monte Carlo steps (MCS).}
\label{fig4}
\end{figure}

According to Mullins' theory, for steps of length
$L$ with periodic boundary conditions, each step configuration, as it
occurs at reference time $\tau_m$,  will eventually decay exponentially
in time, with the relaxation
time increasing like $L^4$ (in the s-d case), and it will
take on a sinusoidal shape with the wavelength
$L$ \cite{mullins}. The correlation functions, $C(x,t)$
and $C_s(x,t)$, indeed, show just this behaviour.
The time needed to reach the asymptotics depends, especially, on the
content of other harmonics in the reference configuration.\\

\noindent {\bf 4. Summary}

Using Monte Carlo techniques, the dynamics of isolated steps and pairs of
steps of monoatomic height has been studied for two types of
kinetics, evaporation--condensation as well as step--edge diffusion,
in the framework of SOS models. Periodic and pinned
boundary conditions are applied.\\

For isolated steps, in addition to the asymptotic behaviour
at late stages, as obtained from
Langevin (or continuum) descriptions, we observe interesting
phenomena at short times, including the superdiffusive step
fluctuations in the case of evaporation kinetics and a non-monotonic time
dependence in the effective exponent in the case of
step--edge diffusion with periodic boundary conditions.\\

For pairs of steps, in the evaporation case, various time scales
are identified where distinct processes govern the step dynamics. 
At late times, applying periodic boundary conditions, the two roughened
steps act like two non--crossing
random walkers. In the case of step--edge diffusion, results
of a recent continuum theory have been checked and extended. In
particular, the time--dependent correlation functions along each
step and between the steps are non--zero, with their long--time
behaviour being described by the classical theory of Mullins on
the flattening of surface corrugations above the roughening
transition temperature.\\ 
 
\noindent {\bf Acknowledgements}
  
We should like to thank M. Bisani, H. P.
Bonzel, P. M. Duxbury, and J. Hager for useful discussions
and information. F. Sz. thanks the
Deutsche Forschungsgemeinschaft, under grant number SE 324/3--1, and
the Hungarian Scientific Research Fund, under grant number OTKA D32835, for 
financial support.\\

\end{document}